\documentclass{amsart} 
\usepackage{graphicx}
\usepackage{sidecap}
\usepackage{amssymb, amsmath}
\usepackage{amsfonts}
\usepackage{amssymb}
\usepackage{float}

\newcommand{\C}{\mathbb C}

\newcommand{\R}{\mathbb R}

\def\la{\label}
\def\tf{\text{ for }}

\def\bt{\begin{thm}}
\def\et{\end{thm}}

\def\bl{\begin{lem}}
\def\el{\end{lem}}

\def\bd{\begin{defi}}
\def\ed{\end{defi}}

\def\bc{\begin{cor}}
\def\ec{\end{cor}}

\def\bp{\begin{proof}}
\def\ep{\end{proof}}

\def\br{\begin{rem}}
\def\er{\end{rem}}

\newtheorem{thm}{Theorem}[section]
\newtheorem{lem}{Lemma}[section]
\newtheorem{defi}{Definition}[section]

\newtheorem{rem}{Remark}[section]
\newtheorem{cor}{Corollary}[section]

\numberwithin{equation}{section}
\numberwithin{theorem}{section}
\numberwithin{example}{section}

\numberwithin{figure}{section}

\begin{document}

\title{Tropical Atmospheric Circulations: Dynamic Stability and Transitions}
\author[Ma]{Tian Ma}
\address[TM]{Department of Mathematics, Sichuan University,
Chengdu, P. R. China}

\author[Wang]{Shouhong Wang}
\address[SW]{Department of Mathematics,
Indiana University, Bloomington, IN 47405}
\email{showang@indiana.edu}

\thanks{The work was supported in part by the
Office of Naval Research and by the National Science Foundation.}


\begin{abstract} In this article, we present a mathematical theory of the Walker circulation of the large-scale atmosphere over the tropics. This study leads to a new metastable state oscillation theory for the El Ni\~no Southern Oscillation (ENSO), a typical inter-annual climate low frequency oscillation. The mathematical analysis is based on 1) the  dynamic transition theory,  2) the geometric theory of incompressible flows, and  3) the scaling law for proper effect of the turbulent friction terms,  developed recently by the authors. 
\end{abstract}

\maketitle
\section{Introduction}
\label{sc1}
The atmosphere and ocean around the earth are rotating geophysical fluids, which are also two important components of the climate system.  The phenomena of the atmosphere and ocean are extremely rich in their organization and complexity, involving a broad range of  temporal and spatial scales.
According to J. von Neumann \cite{neumann}, the motion of the atmosphere can be divided  into three categories depending on the time scale of the prediction. They are motions corresponding respectively to the short time, medium range and long term  behavior of the atmosphere.
One of the primary goals in climate dynamics is to document, through careful 
theoretical and numerical  studies, 
the presence of climate low frequency variability, 
to verify the robustness of this variability's characteristics to 
changes in model parameters, and to help explain its physical mechanisms. 
The thorough understanding of this variability is a 
challenging problem with important practical implications 
for geophysical efforts to quantify predictability, analyze 
error growth in dynamical models,
and develop efficient forecast methods.
One typical source of the inter-annual climate low frequency variability is 
El Ni\~no-Southern Oscillation (ENSO), which is associated with the Walker circulation of the atmosphere over the tropics. The main objective of this article is to  introduce  a mathematical theory of the general circulation of the atmosphere over the tropics.

The modeling part of the study is based on two important ingredients. First, from the general circulation point of view, there are three invariant regions of the atmosphere: the northern hemisphere, the southern hemisphere and the region over the tropics corresponding to zero latitude. The second is the new scaling law introduced recently by the authors, leading to correct circulation length scales. 

The model is then analyzed using the dynamic transition theory and the geometric theory of incompressible flows, both developed recently by the authors. We refer the interested readers to \cite{amsbook} and the references therein for the geometric theory, and to the appendix, \cite{b-book, ptd}  and the references therein  for the dynamic transition theory. We mention in particular here that the main philosophy of the dynamic transition theory is to search for  the full set of  transition states, giving a complete characterization on stability and  transition. This complete set of transition  states represent the {\bf ``physical reality"} after the transition, and  are described by a  local attractor, rather than some steady states or periodic solutions or other type of orbits as part of this local attractor. With this theory, many longstanding phase transition problems are either solved or become more accessible leading to a number of physical predictions. 
For example, the study of phase transitions of  helium-3  leads not only to a theoretical understanding of the phase transitions to superfluidity observed by experiments, but also to such  physical predictions as the  existence of  a new superfluid phase C for liquid helium-3 \cite{MW08g}. 

The main results obtained in this article, on the one hand, verifies the general circulation patterns over the tropics associated with the Walker circulation, and, on the other hand, lead to a new mechanism of El 
Ni\~no Southern Oscillation (ENSO), which will be published in an accompanying paper \cite{MW08l}. 
 This  new mechanism of the ENSO amounts to saying that ENSO, as a self-organizing and self-excitation system, with  two highly coupled processes. The first is the oscillation between the two metastable  warm (El Ni\~no phase)  and cold events (La Ni\~na phase), and the second is the spatiotemporal oscillation of the  sea surface temperature (SST) field. 
The interplay between these two processes gives rises the climate variability associated with the ENSO, leads to both the random and deterministic features of the ENSO,  and 
defines a new natural feedback mechanism, which drives the sporadic oscillation of the ENSO.
 
The paper is organized as follows. In Section 2, we introduce the atmospheric circulation model over the tropics, which is analyzed in Sections 3 and 4, with concluding remarks given in Section 5.

\section{Atmospheric Model over the Tropics}
Physical laws governing the motion and states of the 
atmosphere and ocean can be described by  the general equations of  
hydrodynamics and thermodynamics. 
As discussed in the Introduction, the atmospheric motion equations over the tropics  are the Boussinesq equations restricted on
$\theta =0$, where the meridional  velocity component
$u_{\theta}$ is set to zero, and the effect of the  turbulent friction is taking into considering using the scaling law derived in \cite{ptd}:
\begin{equation}
\begin{aligned}
&\frac{\partial u_{\varphi}}{\partial t}=-(u\cdot\nabla
)u_{\varphi}-\frac{u_{\varphi}u_z}{a}+\nu\left(\Delta
u_{\varphi}+\frac{2}{a^2 }\frac{\partial
u_z}{\partial\varphi}-\frac{2u_{\varphi}}{a^2}\right)\\
&\ \ \ \ -\sigma_0u_{\varphi}-2\Omega
u_z-\frac{1}{\rho_0a}\frac{\partial p}{\partial\varphi},\\
&\frac{\partial u_z}{\partial t}=-(u\cdot\nabla
)u_z+\frac{u^2_{\varphi}}{a}+\nu\left(\Delta
u_z-\frac{2}{a^2 }\frac{\partial
u_{\varphi}}{\partial\varphi}-\frac{2u_z}{a^2 }\right)\\
&\ \ \ \ -\sigma_1u_z+2\Omega
u_{\varphi}-\frac{1}{\rho_0}\frac{\partial p}{\partial
z}-(1-\alpha_T (T-T_0))g,\\
&\frac{\partial T}{\partial t}=-(u\cdot\nabla )T+\kappa\Delta T + Q,\\
&\frac{1}{a }\frac{\partial
u_{\varphi}}{\partial\varphi}+\frac{\partial u_z}{\partial z}=0.
\end{aligned}
\label{10.38}
\end{equation}
Here  $\sigma_i=C_i h^2$ ($i=0, 1$)  represent the turbulent friction,  $a$  is the radius of the  earth, the space domain is taken as $M=S^1_a\times (a ,a +h)$  with $S^1_a$ being the one-dimensional circle with radius $a$, and 
$$
(u\cdot\nabla )=\frac{u_{\varphi}}{a }\frac{\partial}{\partial\varphi}+u_z\frac{\partial}{\partial
z},\qquad 
\Delta =\frac{1}{a^2 }\frac{\partial^2}{\partial\varphi^2}+\frac{\partial^2}{\partial
z^2}.
$$

For simplicity, we denote
$$ (x_1,x_2)=(a \varphi ,z),\qquad (u_1,u_2)=(u_{\varphi},u_z).
$$
In atmospheric physics, the temperature $T_1$ at the tropopause
$z=a +h$ is a constant. We take $T_0$ as the average on the lower
surface $z=a $. To make the nondimensional form, let
\begin{align*}
&x=hx^{\prime},  \qquad t=h^2t^{\prime}/\kappa ,\qquad u=\kappa u^{\prime}/h, \\
&T=(T_0-T_1)T^{\prime}+T_0-(T-T_0)x^{\prime}_2,\\
&p=\kappa\nu\rho_0p^{\prime}/h^2-g\rho_0(hx^{\prime}_2+\alpha_T (T_0-T_1)hx^{\prime
2}/2).
\end{align*}
Also, we define the Rayleigh number, the Prandtl number and the scaling laws by 
\begin{equation}\label{scalinglaw}
R=\frac{\alpha_T g(T_0-T_1)h^3}{\kappa \nu},  \qquad \text{\rm Pr }=\frac{\nu}{\kappa}, \qquad 
 \delta_i=C_ih^4/\nu \quad (i=0, 1).
\end{equation}
Omitting the primes, the nondimensional form of (\ref{10.38}) reads
\begin{equation}
\begin{aligned}
&\frac{\partial u_1}{\partial t}=\text{Pr }\left[\Delta
u_1+\frac{2}{r_0}\frac{\partial u_2}{\partial
x_1}-\frac{2}{r_0}u_1-\delta_0u_1-\frac{\partial p}{\partial
x_1}\right]\\
&\ \ \ \ -\omega u_2-(u\cdot\nabla )u_1-\frac{1}{r_0}u_1u_2,\\
&\frac{\partial u_2}{\partial t}=\text{Pr }\left[\Delta
u_2-\frac{2}{r_0}\frac{\partial u_1}{\partial
x_1}-\frac{2}{r_0}u_2-\delta_1u_2+RT-\frac{\partial p}{\partial
x_2}\right]\\
&\ \ \ \ +\omega u_1-(u\cdot\nabla )u_2-\frac{1}{r_0}u^2_1,\\
&\frac{\partial T}{\partial t}=\Delta T+u_2-(u\cdot\nabla )T +Q,\\
&\frac{\partial u_1}{\partial x_1}+\frac{\partial u_2}{\partial
x_2}=0,
\end{aligned}
\label{10.39}
\end{equation}
where $(x_1,x_2)\in M=(0,2\pi r_0)\times (r_0,r_0+1)$, $\delta_0$ and
$\delta_1$ are as in (\ref{scalinglaw}), $(u\cdot\nabla )$ and $\Delta$
as usual differential operators, and
\begin{equation}
\omega =\frac{2\Omega h^2}{\kappa}.\label{10.40}
\end{equation}

The problem is supplemented with the natural periodic boundary condition in the  $x_1$-direction, and the free-slip boundary condition on the top and bottom boundary:
\begin{align}
& (u, T)(x_1+2\pi r_0,x_2, t)=(u, T)(x_1,x_2, t),  \label{10.41}\\
& 
\left\{
\begin{aligned} 
&u_2=0,\ \frac{\partial u_1}{\partial
x_2}=0,  T=\varphi (x_1) \qquad   && \text{at}\ x_2=r_0,\\
&u_2=0,\ \frac{\partial u_1}{\partial
x_2}=0,\ T=0\qquad  &&  \text{at}\ x_2=r_0+1.
\end{aligned}
\right. \label{10.44}
\end{align}
Here $\varphi (x_1)$ is the temperature deviation from the average
$T_0$ on the equatorial surface and is periodic, i.e.,
$$\int^{2\pi r_0}_0\varphi (x_1)dx_1=0\ \ \ \ \text{and  }\ \varphi
(x_1)=\varphi (x_1+2\pi r_0).$$ The
deviation $\varphi (x_1)$ is mainly caused by a difference in the
specific heat capacities between the sea water and land.

\section{Walker Circulation under the Idealized Conditions}
In an idealized case, the temperature deviation
$\varphi$ vanishes. In this case, the study of transition of
(\ref{10.39}) is of special importance to understand the
longitudinal circulation. Here, we are devoted to discuss the
dynamic bifurcation of (\ref{10.39}), the Walker cell structure of
bifurcated solutions, and the convection scale under the idealized
boundary condition
\begin{equation}
\begin{aligned}
& \varphi (x_1)=0 && 
\text{ for any }  0\leq x_1\leq 2\pi r_0, \\
& Q=0 && \text{ in (\ref{10.39})}.
\end{aligned}
\label{10.45}
\end{equation}

For the problem (\ref{10.39}) with  (\ref{10.41}) and (\ref{10.44}), let 
\begin{eqnarray*}
&&H=\{(u,T)\in L^2(M)^3 \ |\ \text{div} u=0, (u,T)\ \text{satisfies}\
(\ref{10.41}\},\\
&&H_1=\{(u,T)\in H^2(M)^3 \cap H\ |\ (u,T)\ \text{satisfies\ one\ of}\
(\ref{10.41})  \text{ and  } (\ref{10.44})\}.
\end{eqnarray*}
Then, define the operators $L_R=A+B_R$ and $G:H_1\rightarrow H$ by
\begin{equation}
\begin{aligned}
&A\psi =P(P_k(\Delta u_1+\frac{2}{r_0}\frac{\partial u_2}{\partial
x_1}-\delta^{\prime}_0u_1),\text{Pr }(\Delta
u_2-\frac{2}{r_0}\frac{\partial u_1}{\partial
x_1}-\delta^{\prime}_1u_2),\Delta T),\\
&B_R\psi =P(-\omega u_2,\omega u_1+\text{Pr }RT,u_2),\\
&G\psi =P(((u\cdot\nabla )u_1+\frac{u_1u_2}{r_0}),((u\cdot\nabla
)u_2-\frac{u^2_1}{r_0}),(u\cdot\nabla )T),
\end{aligned}
\label{10.46}
\end{equation}
where $\psi =(u,T)\in H_1$,  $P:L^2(M)^3 \rightarrow H$ is the Leray
Projection, and $\delta^{\prime}_i=\frac{2}{r^2_0}+\delta_i$  $(i=0,1)$.
Under the definitions (\ref{10.46}), the problem (\ref{10.39})  with one of (\ref{10.41})  and (\ref{10.44}) is equivalent
to the following abstract equation
\begin{equation}
\frac{d\psi}{dt}=L_R\psi +G(\psi ).\label{10.47}
\end{equation}

Consider the eigenvalue problem
\begin{equation}
L_R\psi =\beta (R)\psi, \label{10.48}
\end{equation}
which is equivalent to 
\begin{equation}
\begin{aligned}
&\Delta u_1+\frac{2}{r_0}\frac{\partial u_2}{\partial
x_1}-\delta^{\prime}_0u_1-P^{-1}_r\omega u_1-\frac{\partial
p}{\partial x_1}=P^{-1}_r\beta u_1,\\
&\Delta u_2-\frac{2}{r_0}\frac{\partial u_1}{\partial
x_1}-\delta^{\prime}_1u_2+P^{-1}_r\omega u_2+RT-\frac{\partial
p}{\partial x_2}=P^{-1}_r\beta u_2,\\
&\Delta T+u_2=\beta T,\\
&\text{div} u=0,
\end{aligned}
\label{10.49}
\end{equation}
with the boundary conditions (\ref{10.41})    and (\ref{10.44}).

We shall see later that these equations (\ref{10.49}) are symmetric, which implies, in particular,  that 
all eigenvalues $\beta_j(R)$ are real, and there
exists a number $R_c$, called the first critical Rayleigh number, such
that
\begin{align}
&
\beta_i(R)\left\{
\begin{aligned} 
&<0  && \text{ if }  R<R_c,\\
&=0 && \text{ if } R=R_c,\\
&>0 && \text{ if } R>R_c
\end{aligned}
\right.  && \text{ for } i=1,2,\label{10.50}
\\
& 
\beta_j(R_c)<0 && \text{ for }  j>2.\label{10.51}
\end{align}

The following theorem provides a theoretical basis to  understand  
the equatorial Walker circulation.

\bt\label{t10.1} Under the idealized condition (\ref{10.45}), the problem (\ref{10.39}) with  (\ref{10.41})   and (\ref{10.44}) undergoes  a Type-I transition  at the critical Rayleigh number $R=R_c$. 
More precisely, the following statements hold true:

\begin{itemize}

\item[(1)] When the Rayleigh number $R\leq R_c$, the equilibrium
solution $(u,T)=0$ is globally stable in the phase space $H$.

\item[(2)] When $R_c<R<R_c+\delta$ for some $\delta >0$, this
problem bifurcates from $((u,T),R)=(0,R_c)$ to an attractor
${\mathcal{A}}_R=S^1$, consisting of steady state
solutions, which attracts $H \setminus \Gamma$ ,   where $\Gamma$ is the stable manifold
of $(u, T)=0$ with codimension two.

\item[(3)] For each steady state solution
$\psi_R=(u_R,T_R)\in{\mathcal{A}_R},u_R$ is topologically equivalent
to the structure as shown in Figure \ref{f10.4}.

\item[(4)] For any initial value $\psi_0=(u_0,T_0)\in H\setminus (\Gamma\cup
E)$, there exists a time $t_0\geq 0$ such that for any $t>t_0$ the
velocity field $u(t,\psi_0)$ is topologically equivalent to the
structure as shown in Figure \ref{f10.5} either (a) or (b), where $\psi
=(u(t,\psi_0),T(t,\psi_0))$ is the solutions of the problem with
$\psi (0)=\psi_0$, and
$$E=\{(u,T)\in H|\ \int^{r_0+1}_{r_0}u_1dx_2=0\}.$$
\end{itemize}
\et
\begin{figure}
  \centering
  \includegraphics[width=0.6\textwidth]{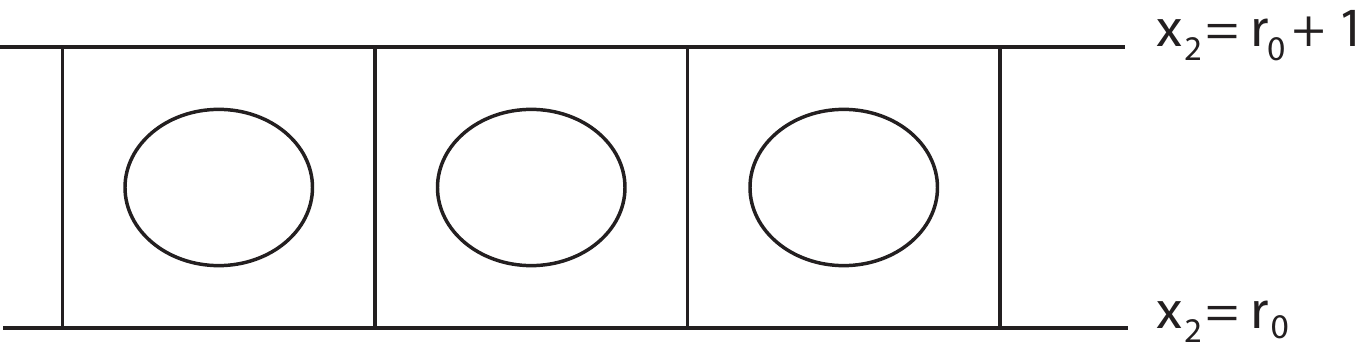}
  \caption{The cell structure of the steady state solutions in the
bifurcated attractor ${\mathcal{A}}_R$.}\la{f10.4}
 \end{figure}

\begin{figure}[hbt]
  \centering
  \includegraphics[width=0.49\textwidth]{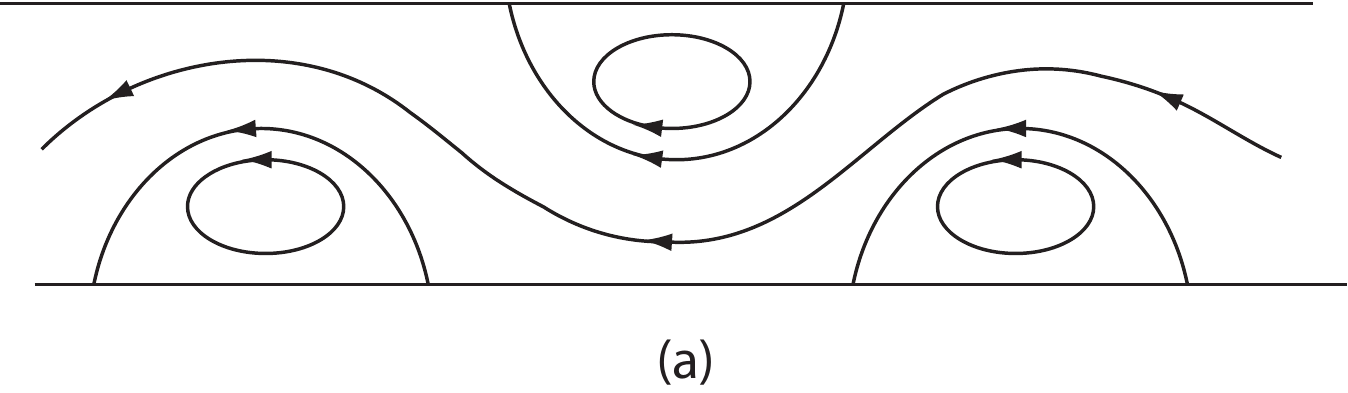} 
  \includegraphics[width=0.49\textwidth]{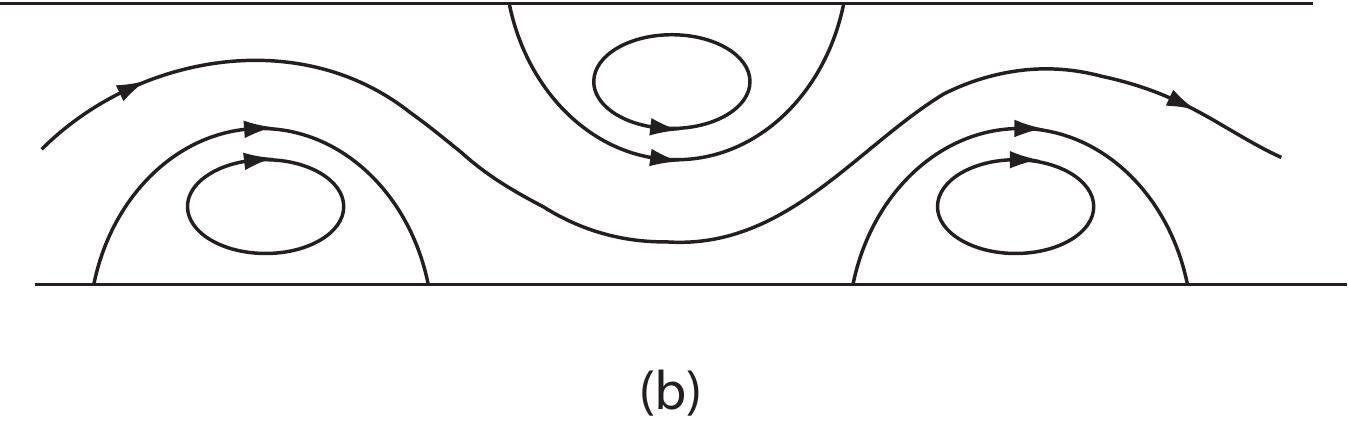}
  \caption{The Walker cell structure with the cells separated by a
cross channel flow: (a) a westbound cross channel flow, and (b) an
eastbound cross channel flow. This cross-channel  flow pattern has the same topological structure as
the Walker circulation over the tropics and the Branstator-Kushnir
waves in the atmospheric dynamics \cite{branstator,kush}.
}\la{f10.5}
 \end{figure}

\bp[Proof of Theorem \ref{t10.1}]  
We proceed in the following several steps.

\medskip

{\sc  Step 1.} We shows that equations (\ref{10.39}) have an equivalent
form as the classical 2D B\'enard problem.

Since the velocity field $u$ defined on $M=S^1\times (r_0,r_0+1)$ is
divergence-free, there exists a stream function $\varphi$ such that
$$u_1=\frac{\partial\varphi}{\partial x_2},\ \ \ \
u_2=-\frac{\partial\varphi}{\partial x_1},$$ 
satisfying  the
given boundary conditions. Therefore, the following  two vector fields
\begin{equation}
\begin{aligned}
&(-\omega u_2,\omega u_1)=\omega\nabla\varphi ,\\
&(\frac{2}{r_0}\frac{\partial u_2}{\partial
x_1},\frac{2}{r_0}\frac{\partial u_1}{\partial
x_1})=-\frac{2}{r_0}\nabla\frac{\partial\varphi}{\partial x_1},
\end{aligned}
\label{10.52}
\end{equation}
are gradient fields, which can be balanced by $\nabla p$ in
(\ref{10.39}). Hence, (\ref{10.39}) are equivalent to 
\begin{equation}
\begin{aligned}
&\frac{\partial u_1}{\partial t}+(u\cdot\nabla
)u_1+\frac{u_1u_2}{r_0}=\text{Pr }\left[\Delta
u_1-\delta^{\prime}_0u_1-\frac{\partial p}{\partial x_1}\right],\\
&\frac{\partial u_2}{\partial t}+(u\cdot\nabla
)u_2-\frac{u^2_1}{r_0}=\text{Pr }\left[\Delta
u_2-\delta^{\prime}_1u_2+RT-\frac{\partial p}{\partial
x_2}\right],\\
&\frac{\partial T}{\partial t}+(u\cdot\nabla )T=\Delta T+u_2,\\
&\text{div} u=0.
\end{aligned}
\label{10.53}
\end{equation}

Therefore, equation (\ref{10.47}) is also an abstract form of
(\ref{10.53}). Thus, equations (\ref{10.48}) and (\ref{10.49}) are
symmetric. It is clear that the operator $G$ defined in
(\ref{10.46}) is orthogonal. Hence, the attractor bifurcation
theorem     
for the Rayleigh-B\'enard convection proved in \cite{MW04d} 
is also valid for the problem (\ref{10.39}) with (\ref{10.41}),
(\ref{10.44}). Thus, this transition of the problem is Type-I (i.e., continuous),
and Assertion (1) is proved.

\medskip

{\sc  Step 2.  Proof of Assertion (2)}. We only need to verify that the
attractor ${\mathcal{A}}_R$ is homeomorphic to a circle, and
consisting of singular points of (\ref{10.47}). 
We consider the first eigenvectors
of (\ref{10.49}) at $R=R_c$. By (\ref{10.52}), equations
(\ref{10.49}) at $R=R_c$ are equivalent to the form
\begin{equation}
\begin{aligned}
&\Delta u_1-\delta^{\prime}_0u_1-\frac{\partial p}{\partial
x_1}=0,\\
&\Delta u_2-\delta^{\prime}_1u_2+R_cT-\frac{\partial p}{\partial
x_2}=0,\\
&\Delta T+u_2=0,\\
&\text{div} u=0,
\end{aligned}
\label{10.54}
\end{equation}
with the boundary conditions (\ref{10.41}) and (\ref{10.44})

As in \cite{MW04d,MW07a}, we see that the multiplicity of $R_c$ in (\ref{10.54}) is $m=2$, and the
corresponding eigenvectors are given by
\begin{align}
& \psi_1=(u_1, u_2, T)=
\left(-\sin\frac{kx_1}{r_0}H^{\prime}(x_2),\frac{k}{r_0}\cos\frac{kx_1}{r_0}H(x_2),\cos\frac{kx_1}{r_0}\Phi (x_2)\right),
\label{10.55}
\\
& 
\tilde{\psi}_1=\left(\tilde{u}_1, \tilde{u}_2, \tilde{T}) = 
( \cos\frac{kx_1}{r_0}H^{\prime}(x_2),\frac{k}{r_0}\sin\frac{kx_1}{r_0}H(x_2),
\sin\frac{kx_1}{r_0}\Phi (x_2)
\right),
\label{10.56}
\end{align}
where   the functions $H(x_2)$ and $\Phi (x_2)$ satisfy that
\begin{align}
& \left\{
\begin{aligned}
&(D^4-(\alpha^2+\alpha^2_0)D^2+\alpha^2\alpha^2_1)H=\alpha R_c\Phi,\\
&-(D^2-\alpha^2)\Phi =\alpha H,
\end{aligned}
\right. 
\label{10.57}
\\
& 
H=H^{\prime\prime}=0,\ \ \ \ \Phi =0 \qquad   \text{ at }  x_2=r_0, r_0+1,\label{10.58}
\end{align}
where $D=d/dx_2,\alpha
=k/r_0,\alpha^2_i=\alpha^2+\delta^{\prime}_i$   $(i=0,1)$.

Since (\ref{10.57}) with (\ref{10.58}) are symmetric, 
we define a number $\alpha$   by 
\begin{equation}
\alpha =\frac{1}{\|\psi_1\|^2}<G(\Psi ,\psi_1)+G(\psi_1,\Psi
),\psi_1>,\label{10.59}
\end{equation}
where $\alpha$ is the number as in (\ref{5.36}), and $\Psi$  is the center manifold function and by \cite{b-book} it satisfies that 
\begin{equation}
-L_{R_c}\Psi =G(\psi_1,\psi_1) +\text{higher order terms},\label{10.60}
\end{equation}
and $L_R$ is defined by (\ref{10.46}), and $G(\tilde{\psi},\psi )$
defined by
$$G(\tilde{\psi},\psi )=P\left(
(\tilde{u}\cdot\nabla)u_1+\frac{\tilde{u}_1u_2}{r_0}, (\tilde{u}\cdot\nabla
)u_2+\frac{\tilde{u}_1u_1}{r_0},  (\tilde{u}\cdot\nabla )T\right),$$
for any $\tilde{\psi}=(\tilde{u},\tilde{T}),\psi =(u,T)\in H_1$.
Obviously, for any  $\psi ,\tilde{\psi}\in H_1$, we  have 
\begin{align*}
&<G(\tilde{\psi},\psi ),\psi >=0,\\
&<G(\psi ,\tilde{\psi}),\psi >=-<G(\psi ,\psi ),\tilde{\psi}>.
\end{align*}
Hence, we derive from (\ref{10.59}) and (\ref{10.60}) that 
$$
\alpha = -\frac{1}{\|\psi_1\|^2}<G(\psi_1,\psi_1),\Psi > = -\frac{1}{\|\psi_1\|^2}<-L_{R_c}\Psi ,\Psi >.
$$
Due to (\ref{10.52}), $-L_R$ is a symmetric sectorial operator.
Thus, we have
$$\alpha =-\frac{1}{\|\psi_1\|^2}<(-L_{R_c})^{{1}/{2}}\Psi
,(-L_{R_c})^{{1}/{2}}\Psi ><0.$$ 
Then, as in \cite{MW07a},   
Assertion (2) follows.

\medskip

{\sc  Step 3.  Proof of Assertion (3).}  By Step 1  and the attractor bifurcation theorem, 
 each steady state solution $\varphi\in{\mathcal{A}}_R$
can be expressed as
\begin{equation}
\varphi =|\beta_1(R)/\alpha
|^{{1}/{2}}(x\psi_1+y\tilde{\psi}_1)+o(\beta_1(R)),\label{10.61}
\end{equation}
where $\psi_1,\tilde{\psi}_1$ are given by (\ref{10.55}) and
(\ref{10.56}), $x^2+y^2=1$, and $\beta_1(R)$ is the first eigenvalue
of (\ref{10.48}) satisfying (\ref{10.50}).

Let $\varphi =(e,T)$. Then, by (\ref{10.61}) the velocity field $e$
is written as
$$e=\left\{\begin{aligned}
&r(R)\cos\frac{k}{r_0}(x_1+\theta )H^{\prime}(x_2)+o(r(R)),\\
&\frac{k}{r_0}r(R)\sin\frac{k}{r_0}(x_1+\theta )H(x_2)+o(r(R)),
\end{aligned}
\right.$$ 
where $r(R)=|\beta_1(R)/\alpha |^{{1}/{2}},(x,y)=(\cos\theta ,\sin\theta )$.

By (\ref{10.57}) and (\ref{10.58}), $H(x_2)=\sin j\pi (x_2-r_0),j\in
Z$. But, for the first eigenvectors $\psi_1$ and $\tilde{\psi}_1$,
we have
$$H(x_2)=\sin\pi (x_2-r_0).$$
Thus, $e$ is expressed as
\begin{equation}
e=e_0+o(r(R)),\label{10.62}
\end{equation}
\begin{equation}
e_0=\left\{\begin{aligned} &\pi r(R)\cos\frac{k}{r_0}(x_1+\theta
)\cos\pi (x_2-r_0),\\
&\frac{k}{r_0}r(R)\sin\frac{k}{r_0}(x_1+\theta )\sin\pi (x_2-r_0),
\end{aligned}
\right.
\label{10.63}
\end{equation}
It is easy to check that $e_0$ is regular, therefore $e$ is also
regular as $R_c<R<R_c+\delta$ for some $\delta >0$. Obviously, the
vector field $e_0$ is topologically equivalent to the structure as
shown in Figure \ref{f10.4}.

By the structural stability theorem in \cite{amsbook}, 
the vector field
$e_0$ in (\ref{10.63}) is not structurally stable in $H_1$, because
the boundary saddle points on $x_2=r_0$ are connected to saddle
points on $x_2=r_0+1$, a different connected component. However,
$e_0$ is structurally stable in the space
$$\tilde{H}=\{(u,T)\in H|\ \int^{2\pi
r_0}_0\int^{r_0+1}_{r_0}udx=0\}.$$ To see this, we know that if
$u\in\tilde{H}$, then $u=(u_1,u_2)\neq$constant, and $u_1$ has the
Fourier expansion
$$u_1=\sum\limits_{k,j}(x_{kj}\cos\frac{kx_1}{r_0}+y_{kj}\sin\frac{kx_1}{r_0})\cos
j\pi (x_2-r_0).$$ It follows that
\begin{equation}
\int^{r_0+1}_{r_0}u_1dx_2=0\ \ \ \ \forall
u=(u_1,u_2)\in\tilde{H}\label{10.64}
\end{equation}
According to the Connection Lemma (Lemma 2.3.1 in Ma and Wang~\cite{amsbook}), from (\ref{10.64}) one can infer that a vector field
$u\in\tilde{H}$ is structurally stable if and only if $u$ is
regular, and all interior saddle points of $u$ are self-connected.
Therefore, $e_0$ is structurally stable in $\tilde{H}$.

If we can prove that the vector field $e$ given by (\ref{10.62}) is
in $\tilde{H}$, then, as $R_c<R<R_c+\delta$,  $e$ is topologically
equivalent to $e_0$. Hence, to prove Assertion (3), it suffices to
verify that $e\in\tilde{H}$.

Obviously, $\tilde{H}$ is an invariant space for the operator
$L_R+G$ defined by (\ref{10.46}), and the orthogonal complementary
$\tilde{H}^{\bot}$ of $\tilde{H}$ in $H$ is
$$\tilde{H}^{\bot}=\{(u,T)\in H|\ u=(c,0),T=0,c\in \R^1\}$$
It is readily to prove that all steady state solutions of
(\ref{10.47}) are in $\tilde{H}$. Thus, Assertion (3) is proved.

\medskip

{\sc Step 4.\ Proof of Assertion (4)}. For any initial value
$\psi_0=(u_0,T_0)\in H \setminus \tilde{H} ,\psi_0$ can be written as
\begin{eqnarray*}
&&\psi_0=c_0+\Phi_0,\Phi_0\in\tilde{H},\\
&&c_0=(\alpha_0,0,0)\in\tilde{H}^{\bot},\ \ \ \ \alpha_0\neq 0.
\end{eqnarray*}
Since $G(\psi )\in\tilde{H},\forall\psi\in H$, the solution $\psi
(t,\psi_0)$ of (\ref{10.47}) with $\psi (0,\psi_0)=\psi_0$ must be
in the form
$$\psi (t,\psi_0)=C_0e^{-\delta^{\prime}_0t}+\Phi (t,\psi_0),\ \ \ \
\Phi\in\tilde{H}.$$ In addition, we know that for each $\psi_0\in
H/\Gamma$ there is a steady state solution $\varphi
=(e,T)\in{\mathcal{A}}_R$ such that
$$\lim\limits_{t\rightarrow\infty}\|\psi
(t,\psi_0)-\varphi\|_{C^r}=0\ \ \ \ \text{for\ any}\ r\geq 1.$$ Then,
for any $\psi_0\in H \setminus (\tilde{H}\cup\Gamma )$, it follows that there
exists a time $t_0>0$ such that the velocity field $u(t,\psi_0)$ of
$\psi (t_0,\psi_0)$ is topologically equivalent to the following
vector field for any $t>t_0$
$$\tilde{u}=e+c_0e^{-\delta^{\prime}_0t}=e+(\alpha_0e^{-\delta^{\prime}_0t},0,0),$$
where $e$ is as in (\ref{10.62}).

The vector field $e$ has the roll structure as shown in Figure \ref{f10.4},
and $c_0e^{-\delta^{\prime}_0t}$ is a parallel channel flow. Hence,
it is easy to show that $\tilde{u}=e+c_0e^{-\delta^{\prime}_0t}$ is
topologically equivalent to the structure as shown in Figure \ref{f10.5}
(a) as $\alpha_0<0$, and to the structure as shown in Figure \ref{f10.5}
(b) as $\alpha_0>0$, for $t>t_0$ large. Assertion (4) is proved.
\ep

\section{Walker circulation under natural conditions}
We now return the natural boundary condition
$$\varphi (x_1)\not\equiv 0 \quad  \text{    and } \quad Q\not=0.$$ 
In this case, equations (\ref{10.39})
admits a steady state solution
\begin{equation}
\tilde{\psi}=(V,J),\ \ \ \ (V=(V_1,V_2)).\label{10.72}
\end{equation}
Consider the deviation from this basic state:
$$u\rightarrow u+V,\ \ \ \ T\rightarrow T+J.$$
Then (\ref{10.39}) becomes
\begin{equation}
\begin{aligned}
&\frac{\partial u_1}{\partial t}+(u\cdot\nabla
)u_1+\frac{u_1u_2}{r_0}=\text{Pr }[\Delta
u_1-\delta^{\prime}_0u_1-\frac{\partial p}{\partial x_1}]\\
&\ \ \ \ -(u\cdot\nabla )u_1-(u\cdot\nabla
)V_1-\frac{V_1}{r_0}u_2-\frac{V_2}{r_0}u_1,\\
&\frac{\partial u_2}{\partial t}+(u\cdot\nabla
)u_2-\frac{u^2_1}{r_0}=\text{Pr }[\Delta
u_2-\delta^{\prime}_1u_2+RT-\frac{\partial p}{\partial x_2}]\\
&\ \ \ \ -(u\cdot\nabla )u_2-(u\cdot\nabla
)V_2+\frac{2V_1}{r_0}u_1,\\
&\frac{\partial T}{\partial t}+(u\cdot\nabla )T=\Delta
T+u_2-(u\cdot\nabla )J-(u\cdot\nabla )T,\\
&\text{div} u=0.
\end{aligned}
\label{10.73}
\end{equation}
The boundary conditions are the free-free boundary conditions given by
\begin{equation}
\begin{aligned} 
&(u,T)\ \text{is\ periodic\ in}\
x_1-\text{direction},\\
&T=0,u_2=0,\frac{\partial u_1}{\partial x_2}=0\ \text{at}\
x_2=r_0,r_0+1.
\end{aligned}
\label{10.74}
\end{equation}

Let the operators $L_R=A+B_R$ and $G:H_1\rightarrow H$ be as in
(\ref{10.46}), and ${\mathcal{L}}^{\varepsilon}:H_1\rightarrow H$ be
defined by
\begin{eqnarray*}
{\mathcal{L}}^{\varepsilon}\psi &=&P(-(V\cdot\nabla
)u_1-(u\cdot\nabla )V_1-\frac{V_1}{r_0}u_2-\frac{V_2}{r_0}u_1,\\
&&-(V\cdot\nabla )u_2-(u\cdot\nabla
)V_2+\frac{2V_1}{r_0}u_1,-(u\cdot\nabla )J-(V\cdot\nabla )T).
\end{eqnarray*}
Then, the problem (\ref{10.73}) and (\ref{10.74}) is equivalent to
the abstract form
\begin{equation}
\frac{d\psi}{dt}=L_R\psi +{\mathcal{L}}^{\varepsilon}\psi +G(\psi
).\label{10.75}
\end{equation}

Consider the eigenvalue problem
\begin{equation}
L_R\psi +{\mathcal{L}}^{\varepsilon}\psi =\beta^{\varepsilon}(R)\psi
.\label{10.76}
\end{equation}

It is known that $|\varphi (x_1)|$ and $Q\not=0$ are small, with $\Delta T=T_0-T_1\cong
100^{\circ}C$ as unit. Hence, the steady state solution
$(V,J)$ is also small:
$$\|(V,J)\|_{L^2}=\varepsilon\ll 1.$$
Thus, (\ref{10.73}) is a perturbation equation of (\ref{10.39}).

Since perturbation terms involving $(V, J)$  are  not invariant under the zonal  translation
(in the $x_1$-direction), for general small functions $\varphi
(x_1)\neq 0$, the first eigenvalues of (\ref{10.73}) are (real or complex) simple, and
by the perturbation theorems in \cite{b-book}, all eigenvalues of linearized equation of (\ref{10.73}) satisfy
the  following principle of exchange of stability (PES):
\begin{align*}
&
\text{Re}\beta^{\varepsilon}_i(R)
\left\{\begin{aligned}
&<0&& \tf  R<R^{\varepsilon}_c,\\
&=0&&\tf R=R^{\varepsilon}_c,\\
&>0&&\tf  R>R^{\varepsilon}_c, 
\end{aligned} \right.   && \text{ for any } 1\leq i\leq m,\\
&  \text{Re}\beta^e_j(R^{\varepsilon}_c)<0  &&  \text{ for any } j\geq
m+1,
\end{align*}
where $m=1$ as $\beta^{\varepsilon}_1(R)$ is real, $m=2$ as
$\beta^{\varepsilon}_1(R)$ is complex near $R^{\varepsilon}_c$, and
$R^{\varepsilon}_c$ is the critical Rayleigh number of perturbed
system (\ref{10.73}).


The following two  theorems follow directly from Theorem~\ref{t10.1} and the perturbation theorems in the appendix or in \cite{b-book}.

\bt\la{t10.2}
Let $\beta^{\varepsilon}_1(R)$ near
$R=R^{\varepsilon}_c$ be a real eigenvalue. Then the system
(\ref{10.73}) has a transition at $R=R^{\varepsilon}_c$, which is
either mixed (Type-III) or continuous (Type-I), depending on the
temperature deviation $\varphi (x_1)$. Moreover, we have the
following assertions:

\begin{itemize}
\item[(1)] If the transition is Type-I, then as
$R^{\varepsilon}_c<R<R^{\varepsilon}_c+\delta$ for some $\delta >0$,
the system bifurcates at $R^{\varepsilon}_c$ to exactly two steady
state solutions $\psi_1$ and $\psi_2$ in $H$, which are
attractors. In particular, space $H$ can be decomposed into two open
sets $U_1,U_2$:
$$H=\bar{U}_1+\bar{U}_2,\ \ \ \ U_1\cap U_2=\emptyset ,\ \ \ \ \psi
=0\in\partial U_1\cap\partial U_2,$$ such that $\psi_i\in
U_i$  $(i=1,2)$, and $\psi_i$ attracts $U_i$.

\item[(2)] If the transition is Type-III, then there is a saddle-node
bifurcation at $R=R^*$ with $R^*<R^{\varepsilon}_c$ such that  the following statements hold true:

\begin{itemize}

\item[(a)] if  $R^*<R<R^{\varepsilon}_c+\delta$ with $R\neq R^{\varepsilon}_c$,
the system has two steady state solutions $\psi^+_R$ and $\psi^-_R$
which are attractors, as shown in  Figure~\ref{f10.10}, such that 
$$\psi^+_R = 0 \qquad \text{ for } R^*<R<R^{\varepsilon}_c.$$

\item[(b)]  There is an open set $U\subset H$ with $0\in U$ which
can be decomposed into two disjoint open sets
$\bar{U}=\bar{U}^R_+ +\bar{U}^R_-$ with $\psi^\pm_R \in U^R_\pm$ and
$\psi^\pm_R$ attracts $U^R_\pm$.
\end{itemize}

\item[(3)] For any initial value $\psi_0=(u_0,T_0)\in U^R_-$ for $R > R^\ast$ or 
$\psi_0=(u_0,T_0)\in U^R_+$  for $R > R^\varepsilon_c$, there exists a time $t_0\geq 0$ such that for any $t>t_0$ the
velocity field $u(t,\psi_0)$ is topologically equivalent to the
structure as shown in Figure \ref{f10.5} either (a) or (b), where $\psi
=(u(t,\psi_0),T(t,\psi_0))$ is the solutions of the problem with
$\psi (0)=\psi_0$.
\end{itemize}
\et
\begin{figure}[hbt]
  \centering
 \includegraphics[width=0.7\textwidth]{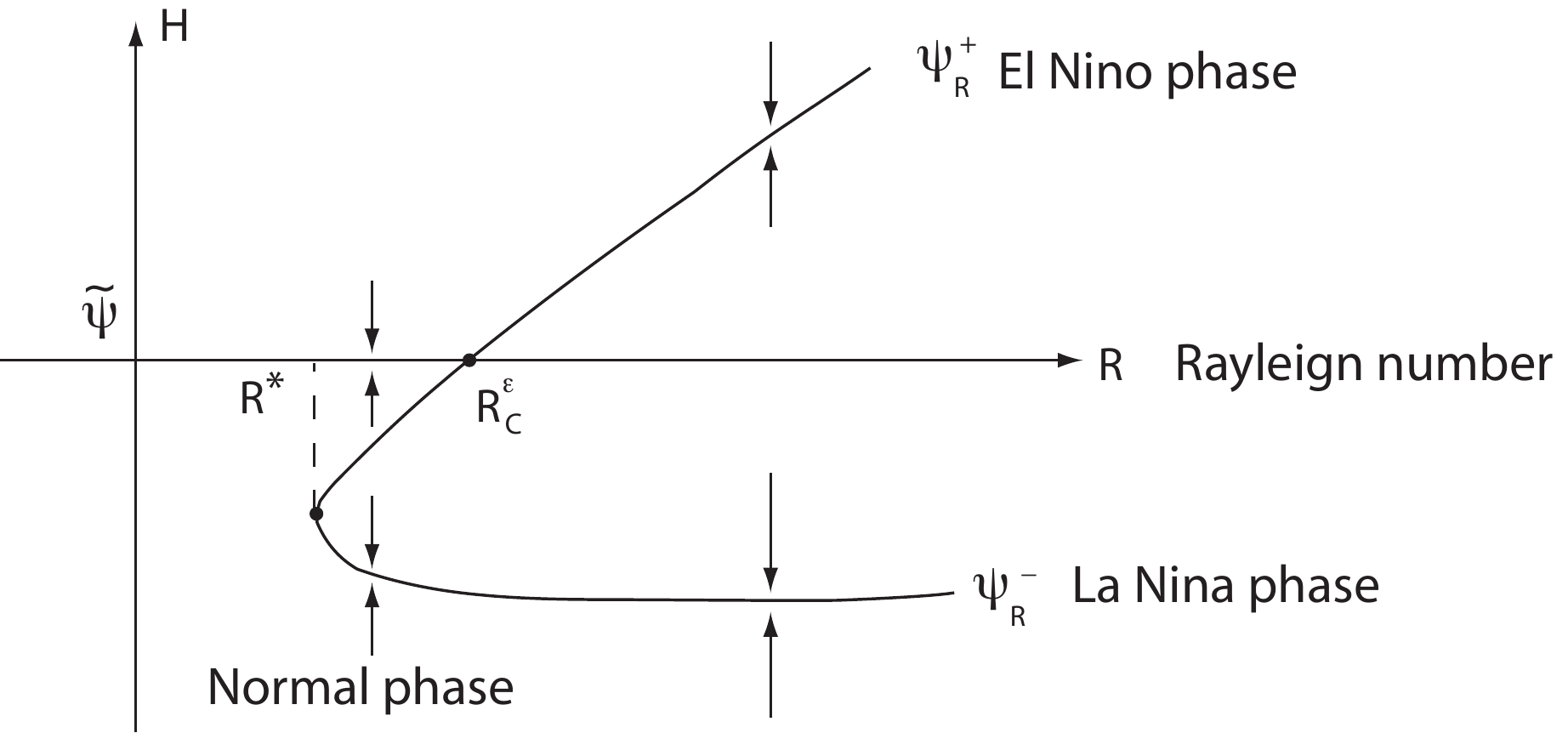}
  \caption{The transition diagram for the equatorial motion
equations (\ref{10.39}), $R^*$ is the saddle-node bifurcation point,
$R^{\varepsilon}_c$ is the first critical Rayleigh number.}\la{f10.10}
 \end{figure}


\bt\la{t10.3}
Let $\beta^{\varepsilon}_1(R)$ be complex near
$R=R^{\varepsilon}_c$. Then the system (\ref{10.73}) bifurcates from
$(\psi ,R)=(0,R^{\varepsilon}_c)$ to a periodic solution $\psi_R(t)$
on $R^{\varepsilon}_c<R$, which is an attractor, and $\psi_R(t)$ can
be expressed as
\begin{equation}
\psi_R(t)=A_R(\cos\rho t\psi_1+\sin\rho
t\tilde{\psi}_1)+o(|A_R|,\varepsilon ),\label{10.77}
\end{equation}
where $A_R=\alpha (R-R^{\varepsilon}_c),\alpha$ and $\rho$ are
constants depending on $\varphi (x_1)$, and $\psi_1,\tilde{\psi}_1$
 are first eigenfunctions of linearized equations of (\ref{10.39}).
\et

\section{Concluding Remarks}
In this article, a careful examination of the dynamic transitions and stability of the large-scale atmospheric flows over the tropics, associated with the Walker circulation and the ENSO,  is given. The analysis and the results obtained show the following from the physical point of view. 

\medskip

{\sc First,} Theorems \ref{t10.1}, \ref{t10.2}   and  \ref{t10.3} provide the possible dynamical behaviors for the  atmospheric  circulation  over the tropics.  Theorem \ref{t10.1} is for the ideal case where the surface temperature profile is given as a constant, leading to  a translation oscillation. Apparently, this result does not represent a realistic explanation to the ENSO.

\medskip

{\sc Second,}  the time-periodic oscillation  obtained in Theorem \ref{t10.3}   
does not represent the typical oscillation in the ENSO phenomena either, as the Walker circulation does not obey the zonal translational oscillation given by the periodic solutions.

\medskip

{\sc Third},  Theorem \ref{t10.2} does give a  characterization of the ENSO, leading  a correct oscillation mechanism of the ENSO between two metastable El Ni\~no and La Ni\~na events, which is further studied in \cite{MW08l}.

\appendix
\setcounter{equation}{0}
\section{Dynamic Transition Theory for Nonlinear Systems}
In this appendix we recall some basic elements of the dynamic transition theory developed by the authors \cite{b-book, chinese-book}, which are used to carry out the dynamic transition analysis for the binary systems in this article. 

\subsection{New classification scheme}
Let $X$  and $ X_1$ be two Banach spaces,   and $X_1\subset X$ a compact and
dense inclusion. In this chapter, we always consider the following
nonlinear evolution equations
\begin{equation}
\left. 
\begin{aligned} 
&\frac{du}{dt}=L_{\lambda}u+G(u,\lambda),\\
&u(0)=\varphi ,
\end{aligned}
\right.\label{5.1}
\end{equation}
where $u:[0,\infty )\rightarrow X$ is unknown function,  and 
$\lambda\in \R^1$  is the system parameter.

Assume that $L_{\lambda}:X_1\rightarrow X$ is a parameterized
linear completely continuous field depending continuously on
$\lambda\in \R^1$, which satisfies
\begin{equation}
\left. 
\begin{aligned} 
&L_{\lambda}=-A+B_{\lambda}   && \text{a sectorial operator},\\
&A:X_1\rightarrow X   && \text{a linear homeomorphism},\\
&B_{\lambda}:X_1\rightarrow X&&  \text{a linear compact  operator}.
\end{aligned}
\right.\label{5.2}
\end{equation}
In this case, we can define the fractional order spaces
$X_{\sigma}$ for $\sigma\in \R^1$. Then we also assume that
$G(\cdot ,\lambda ):X_{\alpha}\rightarrow X$ is $C^r(r\geq 1)$
bounded mapping for some $0\leq\alpha <1$, depending continuously
on $\lambda\in \R^1$, and
\begin{equation}
G(u,\lambda )=o(\|u\|_{X_{\alpha}}),\ \ \ \ \forall\lambda\in
\R^1.\label{5.3}
\end{equation}

Hereafter we always assume the conditions (\ref{5.2}) and
(\ref{5.3}), which represent that the system (\ref{5.1}) has
a dissipative structure.

Let the eigenvalues (counting multiplicity) of $L_{\lambda}$ be given by
$$\{\beta_j(\lambda )\in \C\ \   |\ \ j=1,2,\cdots\}$$
Assume that
\begin{align}
&  \text{Re}\ \beta_i(\lambda )
\left\{ 
 \begin{aligned} 
 &  <0 &&    \text{ if } \lambda  <\lambda_0,\\
& =0 &&      \text{ if } \lambda =\lambda_0,\\
& >0&&     \text{ if } \lambda >\lambda_0,
\end{aligned}
\right.   &&  \forall 1\leq i\leq m,  \label{5.4}\\
&\text{Re}\ \beta_j(\lambda_0)<0 &&  \forall j\geq
m+1.\label{5.5}
\end{align}

The following theorem is a basic principle of transitions from
equilibrium states, which provides sufficient conditions and a basic
classification for transitions of nonlinear dissipative systems.
This theorem is a direct consequence of the center manifold
theorems and the stable manifold theorems; we omit the proof.

\bt\la{t5.1}
 Let the conditions (\ref{5.4}) and
(\ref{5.5}) hold true. Then, the system (\ref{5.1}) must have a
transition from $(u,\lambda )=(0,\lambda_0)$, and there is a
neighborhood $U\subset X$ of $u=0$ such that the transition is one
of the following three types:

\begin{itemize}
\item[(1)] {\sc Continuous Transition}: 
there exists an open and dense set
$\widetilde{U}_{\lambda}\subset U$ such that for any
$\varphi\in\widetilde{U}_{\lambda}$,  the solution
$u_{\lambda}(t,\varphi )$ of (\ref{5.1}) satisfies
$$\lim\limits_{\lambda\rightarrow\lambda_0}\limsup_{t\rightarrow\infty}\|u_{\lambda}(t,\varphi
)\|_X=0.$$ In particular, the attractor bifurcation of (\ref{5.1})
at $(0,\lambda_0)$ is a continuous transition.

\item[(2)] {\sc Jump Transition}: 
for any $\lambda_0<\lambda <\lambda_0+\varepsilon$ with some $\varepsilon >0$, there is an open
and dense set $U_{\lambda}\subset U$ such that 
for any $\varphi\in U_{\lambda}$, 
$$\limsup_{t\rightarrow\infty}\|u_{\lambda}(t,\varphi
)\|_X\geq\delta >0,$$ 
where $\delta >0$ is independent of $\lambda$. 
This type of transition  is also called the discontinuous 
transition. 

\item[(3)] {\sc Mixed Transition}: 
for any $\lambda_0<\lambda <\lambda_0+\varepsilon$  with some $\varepsilon >0$, 
$U$ can be decomposed into two open sets
$U^{\lambda}_1$ and $U^{\lambda}_2$  ($U^{\lambda}_i$ not necessarily
connected):
$$\bar{U}=\bar{U}^{\lambda}_1+\bar{U}^{\lambda}_2,\ \ \
\ U^{\lambda}_1\cap U^{\lambda}_2=\emptyset ,$$ 
such that
\begin{align*}
&\lim\limits_{\lambda\rightarrow\lambda_0}\limsup_{t\rightarrow\infty}\|u(t,\varphi
)\|_X=0   &&   \forall\varphi\in U^{\lambda}_1,\\
& \limsup_{t\rightarrow\infty}\|u(t,\varphi
)\|_X\geq\delta >0 && \forall\varphi\in U^{\lambda}_2.
\end{align*}
\end{itemize}
\et

With this theorem in our disposal, we are in position to give a new dynamic classification scheme for dynamic phase transitions.

\begin{defi}[Dynamic Classification of Phase Transition]
The phase transitions for  (\ref{5.1}) at $\lambda =\lambda_0$ is classified using  their  dynamic properties: continuous, jump, and mixed as given in Theorem~\ref{t5.1}, which are called Type-I, Type-II and Type-III respectively.
\end{defi}

An important aspect of the  transition theory is to determine which 
of the three types of transitions given by Theorem \ref{t5.1} occurs in
a specific  problem. Hereafter we present a few theorems in this theory to be used in this article, and  we refer interested readers to \cite{ptd} for a complete description of the theory.

\subsection{Transitions from simple eigenvalues}
We consider the transition of (\ref{5.1}) from a simple critical
eigenvalue. Let the eigenvalues $\beta_j(\lambda )$ of
$L_{\lambda}$ satisfy (\ref{5.4})  and (\ref{5.5}) with $m=1$.
Then the first eigenvalue $\beta_1(\lambda )$ must be a real eigenvalue. 
Let $e_1(\lambda )$ and $e^*_1(\lambda )$   be  the eigenvectors of
$L_{\lambda}$ and $L^*_{\lambda}$ respectively corresponding to
$\beta_1(\lambda )$ with
$$L_{\lambda_0}e_1=0,\ \ \ \ L^*_{\lambda_0}e^*_1=0,\ \ \ \
<e_1,e^*_1>=1.$$ 
Let $\Phi (x,\lambda )$    be the center manifold
function of (\ref{5.1}) near $\lambda =\lambda_0$. We assume that
\begin{equation}
<G(xe_1+\Phi (x,\lambda_0),\lambda_0),e^*_1>=\alpha
x^k+o(|x|^k),\label{5.36}
\end{equation}
where $k\geq 2$ an integer and $\alpha\neq 0$ a real number.
\begin{figure}
  \centering
  \includegraphics[width=0.35\textwidth]{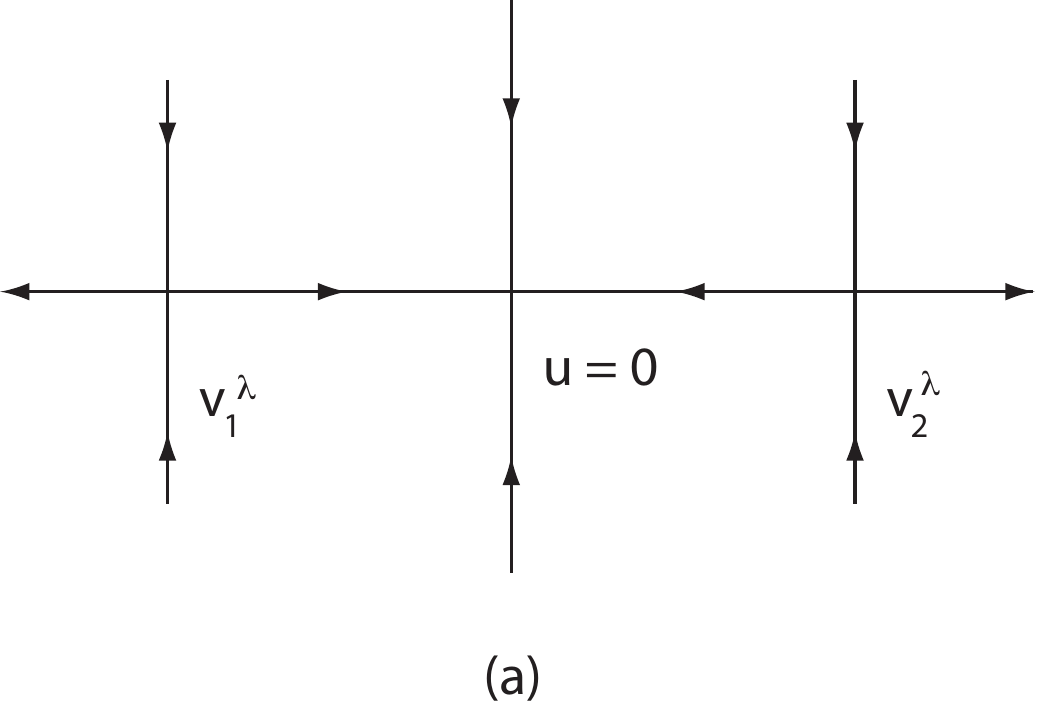} \quad 
  \includegraphics[width=0.2\textwidth]{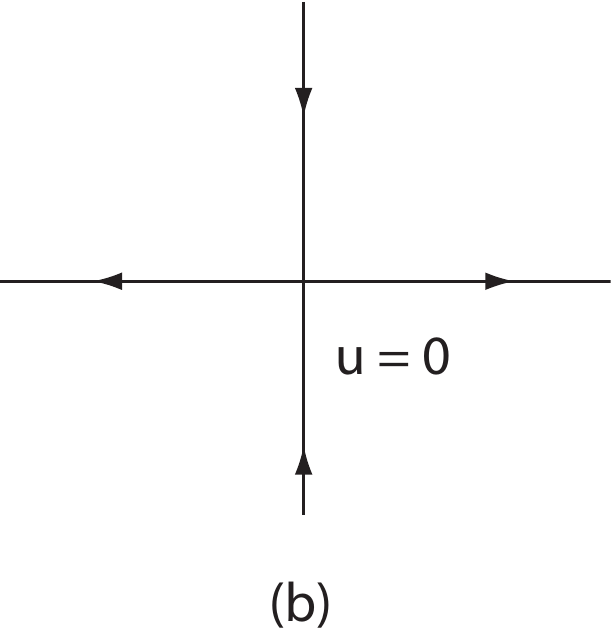}
  \caption{Topological structure of the jump transition of
(\ref{5.1}) when $k$=odd and $\alpha >0$: (a) $\lambda
<\lambda_0$; (b) $\lambda\geq\lambda_0$. Here the horizontal line
represents the center manifold.}\la{f5.5}
 \end{figure}
 \begin{figure}
  \centering
  \includegraphics[width=0.23\textwidth]{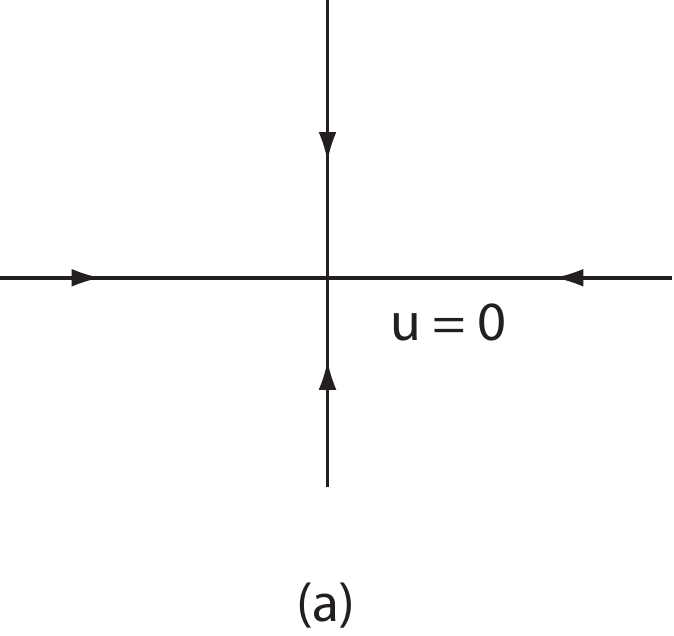}
  \includegraphics[width=0.35\textwidth]{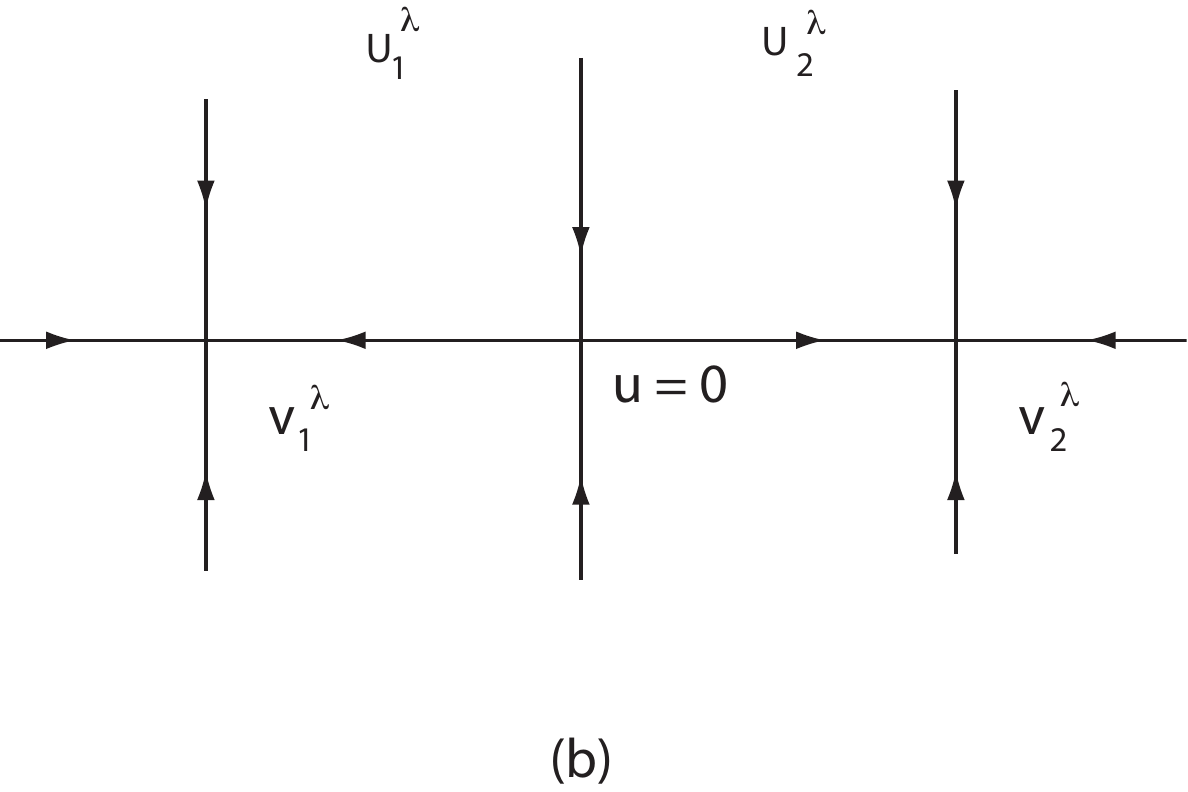}
  \caption{Topological structure of the continuous transition
of (\ref{5.1}) when $k$=odd and $\alpha <0$: (a)
$\lambda\leq\lambda_0$; (b) $\lambda >\lambda_0$.}\la{f5.6}
 \end{figure}
 \begin{figure}
  \centering
  \includegraphics[width=0.32\textwidth]{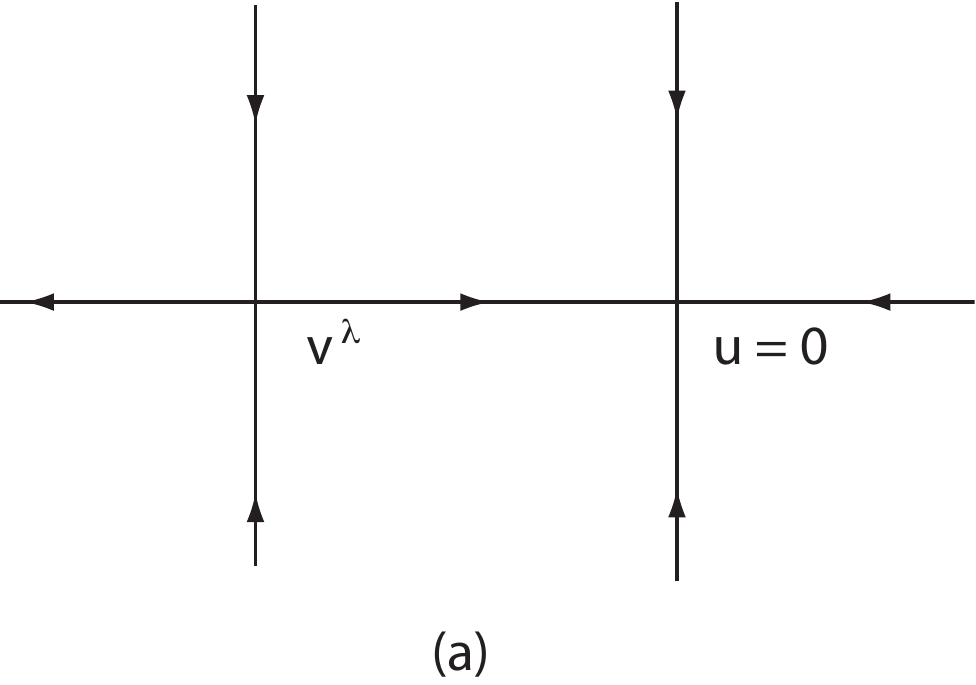}
   \includegraphics[width=0.22\textwidth]{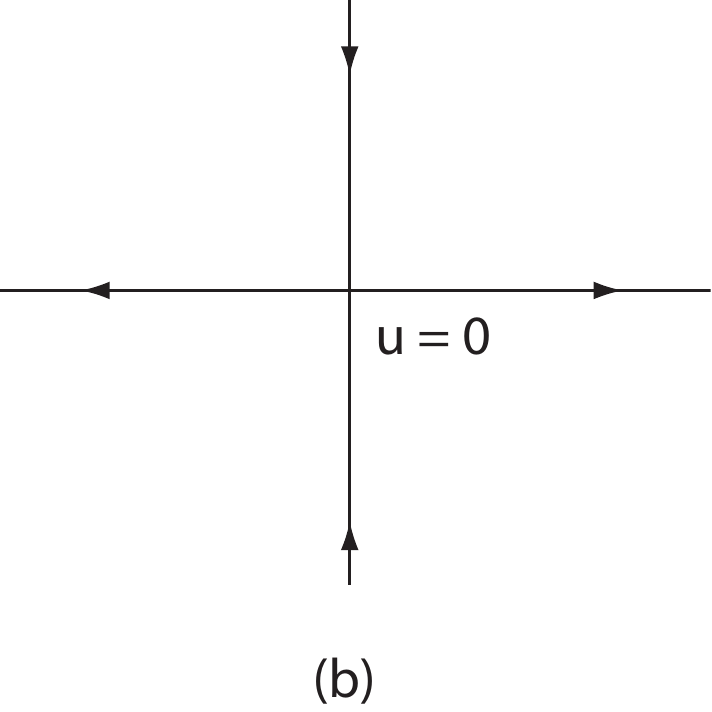} 
   \includegraphics[width=0.32\textwidth]{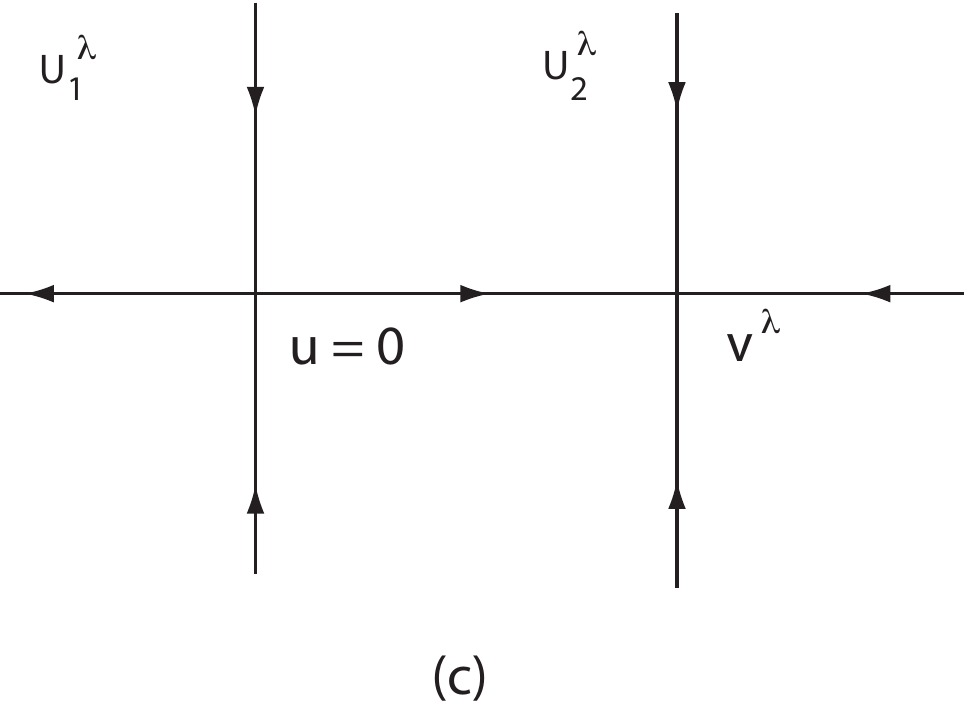}
  \caption{Topological structure of the mixing transition of
(\ref{5.1}) when $k$=even and $\alpha\neq 0$: (a) $\lambda
<\lambda_0$; (b) $\lambda =\lambda_0$; (c) $\lambda >\lambda_0$. Here
$U^{\lambda}_1$ is the unstable domain, and $U^{\lambda}_2$ the
stable domain.}\la{f5.7}
 \end{figure}

\bt\la{t5.8}
Assume  (\ref{5.4})  and (\ref{5.5}) with $m=1$, and (\ref{5.36}).  If $k$=odd and $\alpha\neq 0$ in (\ref{5.36}) then
the following assertions hold true:

\begin{itemize}

\item[(1)] If $\alpha >0$,  then (\ref{5.1}) has a jump
transition from $(0,\lambda_0)$, and bifurcates on $\lambda
<\lambda_0$ to exactly two saddle points $v^{\lambda}_1$ and
$v^{\lambda}_2$ with the Morse index one, as shown in Figure \ref{f5.5}.

\item[(2)] If $\alpha <0$,  then (\ref{5.1}) has a continuous
transition from $(0,\lambda_0)$, which is an attractor bifurcation
 as shown in Figure \ref{f5.6}. 

\item[(3)] The bifurcated singular points $v^{\lambda}_1$ and $v^{\lambda}_2$ 
in the above cases can
be expressed in the following form
$$v^{\lambda}_{1,2}=\pm |\beta_1(\lambda )/\alpha
|^{{1}/{k-1}}e_1(\lambda )+o(|\beta_1|^{{1}/{k-1}}).$$

\end{itemize}
\et

\bt\la{t5.9}
 Assume  (\ref{5.4})  and (\ref{5.5}) with $m=1$,  and
(\ref{5.36}). If $k$=even and $\alpha\neq 0$, then we have the
following assertions:

\begin{enumerate}

\item (\ref{5.1}) has a mixed transition from
$(0,\lambda_0)$. More precisely, there exists a neighborhood
$U\subset X$ of $u=0$ such that $U$ is separated into two disjoint
open sets $U^{\lambda}_1$ and $U^{\lambda}_2$ by the stable
manifold $\Gamma_{\lambda}$ of $u=0$ satisfying the following properties:

\begin{enumerate}

\item $U=U^{\lambda}_1+U^{\lambda}_2+\Gamma_{\lambda}$,

\item the transition in $U^{\lambda}_1$ is jump, and 

\item the transition in $U^{\lambda}_2$ is
continuous. The local transition structure is as shown in Figure \ref{f5.7}.

\end{enumerate}

\item (\ref{5.1}) bifurcates in $U^{\lambda}_2$ to a unique
singular point $v^{\lambda}$ on $\lambda >\lambda_0$, which is an
attractor such that for any $\varphi\in U^{\lambda}_2$, 
$$\lim\limits_{t\rightarrow\infty}\|u(t,\varphi
)-v^{\lambda}\|_X=0,$$
where $u(t,\varphi )$ is the solution of (\ref{5.1}). 

\item (\ref{5.1})\ bifurcates on $\lambda <\lambda_0$ to a unique saddle
point $v^{\lambda}$ with the Morse index one. 

\item The bifurcated singular point $v^{\lambda}$ can be expressed as
$$v^{\lambda}=-(\beta_1(\lambda )/\alpha
)^{{1}/{(k-1)}}e_1+o(|\beta_1|^{{1}/{(k-1)}}).$$
\end{enumerate}
\et

\subsection{Singular Separation}
\la{s6.2}
In this section, we study  an important
problem associated with the discontinuous transition of
(\ref{5.1}), which we call  the singular separation.

\bd\la{d6.1}
\begin{enumerate}

\item 
An invariant set $\Sigma$ of (\ref{5.1}) is called a singular
element if $\Sigma$ is either a singular point or a periodic
orbit. 

\item Let $\Sigma_1\subset X$ be a singular
element of (\ref{5.1}) and $U\subset X$ a neighborhood of
$\Sigma_1$. We say that (\ref{5.1}) has a singular separation of
$\Sigma$ at $\lambda =\lambda_1$ if 

\begin{enumerate}

\item (\ref{5.1}) has no singular
elements in $U$ as $\lambda <\lambda_1$ (or $\lambda >\lambda_1$),
and generates a singular element $\Sigma_1\subset  U$ at $\lambda
=\lambda_1$,  and 

\item there are branches of singular elements
$\Sigma_{\lambda}$, which are  separated from $\Sigma_1$ for $\lambda
>\lambda_1$ (or $\lambda <\lambda_1$), i.e.,
$$\lim\limits_{\lambda\rightarrow\lambda_1}\max_{x\in\Sigma_{\lambda}}\text{dist}(x,\Sigma_1)=0.$$
\end{enumerate}

\end{enumerate}
\ed

A special case of singular separation is the saddle-node
bifurcation defined as follows.

\bd\la{d6.2}
Let $u_1\in X$ be a singular point of
(\ref{5.1}) at $\lambda =\lambda_1$ with $u_1\neq 0$. We say that
(\ref{5.1}) has a saddle-node bifurcation at $(u_1,\lambda_1)$ if

\begin{enumerate}

\item the index of $L_{\lambda}+G$ at $(u_1,\lambda_1)$ is zero, i.e.,
ind$(-(L_{\lambda_1}+G),u_1)=0$, 

\item  there are two branches
$\Gamma_1(\lambda )$ and $\Gamma_2(\lambda )$ of singular points
of (\ref{5.1}), which  are separated from $u_1$ for $\lambda >\lambda_1$
(or $\lambda <\lambda_1)$,  i.e.,  for any
$u_{\lambda}\in\Gamma_i(\lambda )$ $(i=1,2)$ we have
$$u_{\lambda}\rightarrow u_1\ \text{in}\ X\ \ \ \ \text{as}\
\lambda\rightarrow\lambda_1, $$ 
and

\item  the indices of
$u^i_{\lambda}\in\Gamma_i(\lambda )$ are as follows
$$\text{ind}(-(L_{\lambda}+G),u_{\lambda})=
\left\{
\begin{aligned}
&  1  &&  \text{ if } u_{\lambda}\in\Gamma_2(\lambda ),\\
&  -1  &&  \text{ if } u_{\lambda}\in\Gamma_1(\lambda ).
\end{aligned}
\right.$$
\end{enumerate}
\ed

Intuitively, the saddle-node bifurcation is
schematically shown as in Figure~\ref{f6.1}, where the singular points in
$\Gamma_1(\lambda )$ are saddle points and in $\Gamma_2(\lambda )$
are nodes, and the singular separation of periodic orbits is as in
shown Figure~\ref{f6.2}.
\begin{figure}[hbt]
  \centering
  \includegraphics[width=0.35\textwidth]{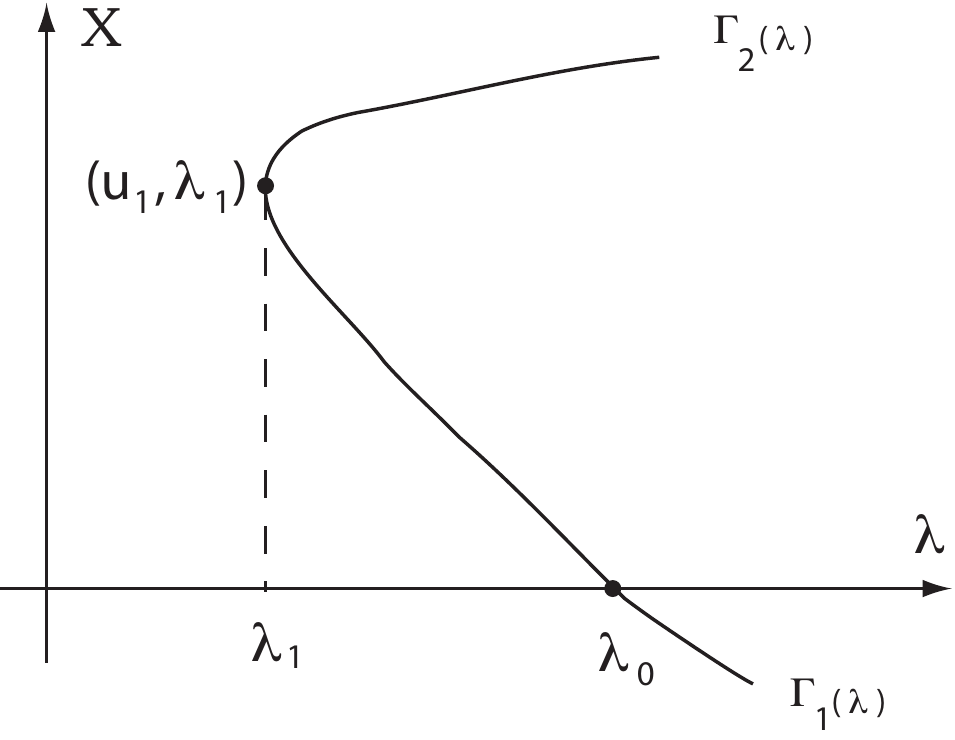}
  \caption{Saddle-node bifurcation.}\la{f6.1}
 \end{figure}
\begin{figure}[hbt]
  \centering
  \includegraphics[width=0.35\textwidth]{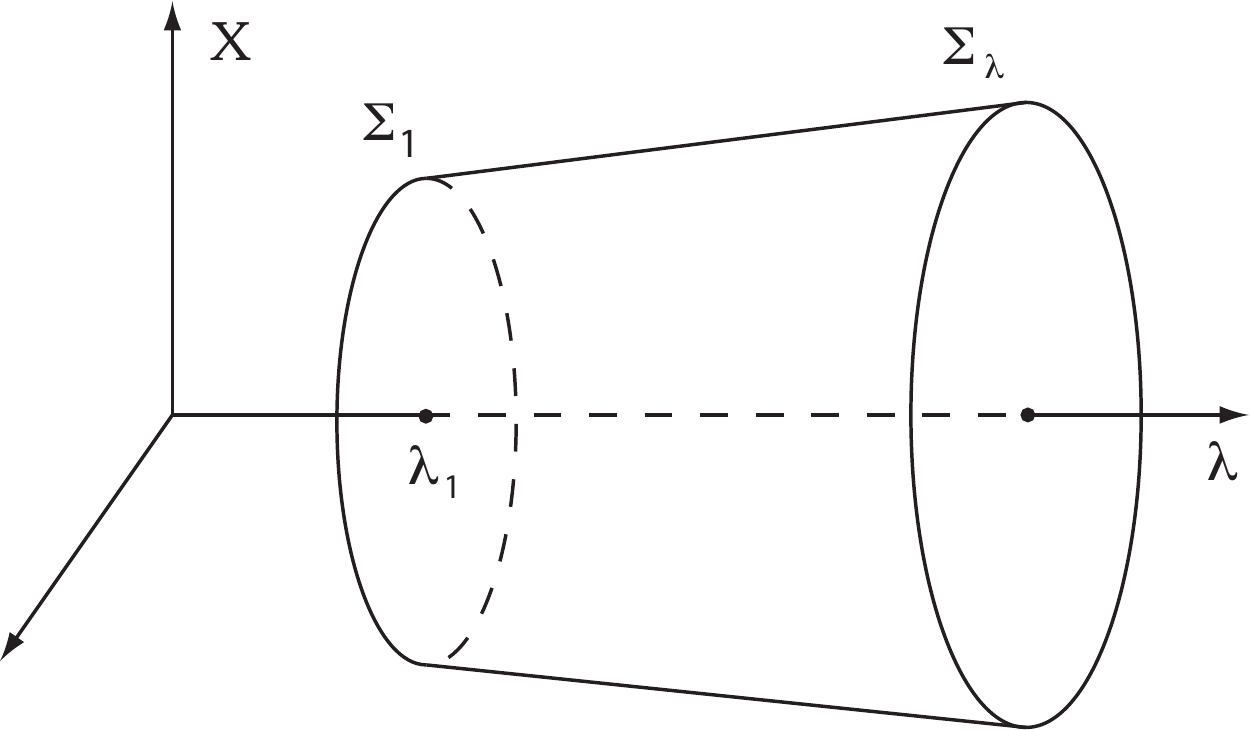}
  \caption{Singular separation of periodic orbits.}\la{f6.2}
 \end{figure}

For the singular separation we can give a general principle as
follows, which provides a basis for singular separation theory.

\bt\la{t6.4}
Let the conditions (\ref{5.4}) and
(\ref{5.5}) hold true. Then we have the following assertions.

\begin{enumerate}

\item[(1)] If (\ref{5.1}) bifurcates from $(u,\lambda
)=(0,\lambda_0)$   to a branch $\Sigma_{\lambda}$ of singular elements on
$\lambda <\lambda_0$ which is bounded in $X\times (-\infty
,\lambda_0)$ then (\ref{5.1}) has a singular separation of
singular elements at some $(\Sigma_0,\lambda_1)\subset X\times
(-\infty ,\lambda_0)$. 

\item[(2)] If the bifurcated branch
$\Sigma_{\lambda}$ consists of singular points which has index $-1$,
i.e., 
$$\text{ind}(-(L_{\lambda}+G),u_{\lambda})=-1 \ \ \ \ \forall u_{\lambda}\in
E_{\lambda},\ \ \ \ \lambda <\lambda_0,$$ then the singular
separation is a saddle-node bifurcation from some
$(u_1,\lambda_1)\in X\times (-\infty ,\lambda_0).$
\end{enumerate}
\et

We consider the equation (\ref{5.1}) defined on the Hilbert spaces
$X=H, X_1=H_1$. Let $L_{\lambda}=-A+\lambda B$. For $L_{\lambda}$
and $G(\cdot ,\lambda ):H_1\rightarrow H$, we assume that
$A:H_1\rightarrow H$ is symmetric, and
\begin{eqnarray}
&&<Au,u>_H\geq c\|u\|^2_{H_{{1}/{2}}},\label{6.46}\\
&&<Bu,u>_H\geq c\|u\|^2_H,\label{6.47}\\
&&<Gu,u>_H\leq -c_1\|u\|^p_H+c_2\|u\|^2_H,\label{6.48}
\end{eqnarray}
where $p>2, c, c_1, c_2>0$ are constants.

\bt\la{t6.5}
 Assume the conditions (\ref{5.3}),
(\ref{5.4}) and (\ref{6.46})-(\ref{6.48}), then (\ref{5.1}) has a
transition at $(u,\lambda )=(0,\lambda_0)$, and the following
assertions hold true:

\begin{enumerate}
\item[(1)] If $u=0$ is an even-order nondegenerate singular point
of $L_{\lambda}+G$ at $\lambda =\lambda_0$, then (\ref{5.1}) has a
singular separation of singular points at some $(u_1,\lambda_1)\in
H\times (-\infty ,\lambda_0)$. 

\item[(2)]  If $m=1$ and $G$
satisfies (\ref{5.36}) with $\alpha >0$ if $k$=odd and $\alpha\neq
0$ if $k$=even, then (\ref{5.1}) has a saddle-node bifurcation at
some singular point $(u_1,\lambda_1)$ with $\lambda_1<\lambda_0$.
\end{enumerate}
\et

\subsection{Transition and Singular Separation of Perturbed Systems}
We consider the following perturbed equation of (\ref{5.1}): 
\begin{equation}
\frac{du}{dt}=(L_{\lambda}+S^{\varepsilon}_{\lambda})u+G(u,\lambda
)+T^{\varepsilon}(u,\lambda ),\label{6.56}
\end{equation}
where $L_{\lambda}$ and $G_{\lambda}$ are as in (\ref{5.1}),
$S^{\varepsilon}_{\lambda}:X_{\sigma}\rightarrow X$ is a linear
perturbed operator,
$T^{\varepsilon}_{\lambda}:X_{\sigma}\rightarrow X$ a $C^1$
nonlinear perturbed operator,  and  $X_{\sigma}$ the fractional order
space, $0\leq\sigma <1$. Also assume that 
$G_{\lambda},T^{\varepsilon}_{\lambda}$ are
$C^3$ on $u$, and
\begin{equation}
\left.
\begin{aligned} 
& \|S^{\varepsilon}_{\lambda}\|<\varepsilon,\\
& \|T^{\varepsilon}_{\lambda}\|<\varepsilon ,\\
& T^{\varepsilon}(u,\lambda )=o(\|u\|_{X_{\alpha}}).
\end{aligned}
\right.\label{6.57}
\end{equation}

Let  (\ref{5.4})  and (\ref{5.5}) with $m=1$ hold true, $G(u,\lambda )=G_2(u,\lambda
)+o(\|u\|^2_{X_1})$, where $G_2(\cdot ,\lambda )$ is a bilinear
operator, and
\begin{equation}
b=<G_2(e,\lambda_0),e^*>\neq 0,\label{6.65}
\end{equation}
where $e\in X$ and $e^*\in X^*$ are the eigenvectors of
$L_{\lambda}$ and $L^*_{\lambda}$ corresponding to
$\beta_1(\lambda )$ at $\lambda =\lambda_0$ respectively.

We now consider the transition associated with the saddle-node
bifurcation of the perturbed system (\ref{6.56}). Let $h(x,\lambda
)$ be the center manifold function of (\ref{5.1}) near $\lambda
=\lambda_0$. Assume that
\begin{equation}
<G(xe+h(x,\lambda_0),\lambda_0),e^*>=b_1x^3+o(|x|^3),\label{6.66}
\end{equation}
where $b_1\neq 0$, and $e$ and $e^*$ are as in (\ref{6.65}).

Then we have the following theorems.

\bt\la{t6.10}
 Let the conditions (\ref{5.4})  and (\ref{5.5}) with $m=1$, and
(\ref{6.66}) hold true, and $b_1<0$. Then there is an
$\varepsilon >0$ such that if  $S^{\varepsilon}_{\lambda}$ and
$T^{\varepsilon}_{\lambda}$ satisfy (\ref{6.57}), then the transition of
(\ref{6.56}) is either continuous or mixed. If the transition is
continuous, then Assertions (2) and (3) of Theorem \ref{t5.8} are valid
for (\ref{6.56}). If the transition is mixed, then the
following assertions hold true:

\begin{enumerate}

\item[(1)] (\ref{6.56}) has a saddle-node bifurcation at some
point $(u_1,\lambda_1)\in X\times (-\infty
,\lambda^{\varepsilon}_0)$, and there are exactly two branches
$$\Gamma^{\lambda}_i=\{(u^{\lambda}_i,\lambda )|\ \lambda_1<\lambda
<\lambda^{\varepsilon}_0+\delta\} \qquad   i=1,2, $$ 
separated from
$(u_1,\lambda_1)$ as shown in Figure~\ref{f6.1}, which satisfy that 
\begin{align*}
&  \|u^{\lambda}_2\|_X\neq 0  &&\forall (u^{\lambda}_2,\lambda
)\in\Gamma^{\lambda}_2,\ \ \ \ \lambda_1<\lambda
<\lambda^{\varepsilon}_0+\delta ,\\
& \lim_{\lambda\rightarrow\lambda^{\varepsilon}_0}\|u^{\lambda}_1\|_X=0 &&
\forall  (u^{\lambda}_1,\lambda )\in\Gamma^{\lambda}_1.
\end{align*}

\item[(2)] There is a neighborhood $U\subset X$ of
$u=0$, such that for each $\lambda$ with $\lambda_1<\lambda
<\lambda^{\varepsilon}_0+\delta$ and
$\lambda\neq\lambda^{\varepsilon}_0, U$ contains only two
nontrivial singular points $u^{\lambda}_1$ and $u^{\lambda}_2$ of
(\ref{6.56}). 

\item[(3)] For each $\lambda_1<\lambda
<\lambda^{\varepsilon}_0+\delta , U$ can be decomposed into two
open sets $\bar{U}=\bar{U}^{\lambda}_1+\bar{U}^{\lambda}_2$ with
$U^{\lambda}_1\cap U^{\lambda}_2=\emptyset$, such that

\begin{enumerate}

\item if $  \lambda_1<\lambda <\lambda^{\varepsilon}_0$, 
$$0\in U^{\lambda}_1,\ \ \ \ u^{\lambda}_2\in U^{\lambda}_2,\ \ \
\ u^{\lambda}_1\in\partial U^{\lambda}_1\cap\partial
U^{\lambda}_2,$$ with $u=0$ and $u^{\lambda}_2$ being
attractors which attract $U^{\lambda}_1$ and $U^{\lambda}_2$
respectively, and

\item if  $ \lambda^{\varepsilon}_0<\lambda <\lambda^{\varepsilon}_0+\delta$,
$$u^{\lambda}_1\in U^{\lambda}_1,\ \ \ \ u^{\lambda}_2\in
U^{\lambda}_2,\ \ \ \ 0\in\partial U^{\lambda}_1\cap\partial
U^{\lambda}_2, $$ 
with $u^{\lambda}_1$ and
$u^{\lambda}_2$ being attractors which attract $U^{\lambda}_1$ and
$U^{\lambda}_2$ respectively. 

\end{enumerate}

\item[(4)] Near $(u,\lambda)=(0,\lambda^{\varepsilon}_0),u^{\lambda}_1$ 
and $u^{\lambda}_2$
can be expressed as 
\begin{equation} 
\left.
\begin{aligned}
& u^{\lambda}_1=\alpha_1(\lambda ,\varepsilon )e+o(|\alpha_1|),\\
& u^{\lambda}_2=\alpha_2(\lambda ,\varepsilon )e+o(|\alpha_2|),\\
& \lim_{  \lambda\rightarrow\lambda^{\varepsilon}_0  }  
\alpha_1(\lambda ,\varepsilon )= 0,\\
& \alpha_2(\lambda^{\varepsilon}_0,\varepsilon )\neq 0,
\end{aligned}
\right.\label{6.67}
\end{equation}
where $e$ is as in (\ref{6.66}).
\end{enumerate}
\et

\bt\la{t6.11} Assume the conditions (\ref{5.4})  and (\ref{5.5}) with $m=1$, and
(\ref{6.66}) with $b_1>0$. Then, there is an $\varepsilon >0$ such
that when $S^{\varepsilon}_{\lambda}$ and
$T^{\varepsilon}_{\lambda}$ satisfy (\ref{6.57}), the transition of
(\ref{6.56}) is either jump or mixed. If it is jump transition,  then
Assertions (1) and  (3) of Theorem \ref{t5.8} are valid for (\ref{6.56}). 
If it is mixed, then the following assertions hold true:

\begin{enumerate}

\item[(1)] (\ref{6.56}) has a saddle-node bifurcation at some
point $(u_1,\lambda_1)\in X\times (\lambda^{\varepsilon}_0,+\infty
)$, and there are exactly two branches
$$\Gamma^{\lambda}_i=\{(u^{\lambda}_i,\lambda )|\
\lambda^{\varepsilon}_0-\delta <\lambda <\lambda_1\} \qquad   (i=1,2), $$
separated from $(u_1,\lambda_1)$, which satisfy
\begin{align*}
&  \|u^{\lambda}_2\|_X=0 &&\forall (u^{\lambda}_2,\lambda)\in\Gamma^*_2,\ \ \ \ \lambda^{\varepsilon}_0-\varepsilon
<\lambda <\lambda_1,\\
& \lim\limits_{\lambda\rightarrow\lambda^{\varepsilon}_0}\|u^{\lambda}_1\|_X=0
&& \forall 
(u^{\lambda}_1,\lambda )\in\Gamma^{\lambda}_1.
\end{align*}

\item[(2)] There is a neighborhood $U\subset X$ of
$u=0$, such that for each $\lambda$ with
$\lambda^{\varepsilon}_0-\delta <\lambda <\lambda_1, U$ contains
only two nontrivial singular points $u^{\lambda}_1$ and
$u^{\lambda}_2$ of (\ref{6.56}).

\item[(3)] For every
$\lambda^{\varepsilon}_0-\delta <\lambda <\lambda_1, U$ can be
decomposed into three open sets
$\bar{U}=\bar{U}_0+\bar{U}_1+\bar{U}_2$ with $U_i\cap
U_j=\emptyset$ $ (i\neq j)$ such that

\begin{enumerate}

\item  if   $ \lambda^{\varepsilon}_0-\delta <\lambda <\lambda^{\varepsilon}_0$, 
then 
$$u=0\in U^{\lambda}_0,\ \ \ \ u^{\lambda}_i\in\partial
U^{\lambda}_i\cap\partial U^{\lambda}_0(i=1,2),$$ 
with $u=0$ being an attractor which
attracts $U^{\lambda}_0$ and $U^{\lambda}_i(i=1,2)$ two saddle
points with the Morse index one, and

\item  if   $ \lambda^{\varepsilon}_0<\lambda <\lambda_1$,  then 
$$u^{\lambda}_1\in U^{\lambda}_1,\ \ \ \ u^{\lambda}_2\in\partial
U^{\lambda}_2\cap\partial U^{\lambda}_1,\ \ \ \ 0\in\partial
U^{\lambda}_1\cap\partial U^{\lambda}_0,$$ 
with $u^{\lambda}_1$
being an attractor which attracts $U^{\lambda}_1$ and
$u^{\lambda}_2$ and $u=0$ being saddle points with the Morse index
one. 

\end{enumerate}

\item[(4)] Near $(0,\lambda^0_{\varepsilon}),u^{\lambda}_1$
and $u^{\lambda}_2$ can be expressed by (\ref{6.67}).
\end{enumerate}
\et

\bibliographystyle{siam}

\end{document}